\documentclass[conference]{IEEEtran}

\ifCLASSINFOpdf
  \usepackage[pdftex]{graphicx}
\else
\fi
\usepackage{amsmath}
\usepackage{url}

\usepackage[]{fancyhdr} %
\newcommand{\changefont}{\fontsize{9}{9}\selectfont}
\usepackage{float}
\IEEEoverridecommandlockouts
\begin{document}

%
\title{Assessment of frequency stability
behaviour\\ regarding inertia reduction
due to high renewable\\ integration in the
Iberian system
}

\author{\IEEEauthorblockN{Francisco Sousa Fernandes, MSc \\
University of Porto }
\IEEEauthorblockA{Faculty of Engineering \\
INESC TEC \\
 Porto, Portugal\\
francisco.s.fernandes@inesctec.pt}

\and

\IEEEauthorblockN{João Peças Lopes, PhD, FIEEE \\
University of Porto }
\IEEEauthorblockA{Faculty of Engineering \\
INESC TEC \\
 Porto, Portugal\\
jpl@fe.up.pt}
}


%





\maketitle
\thispagestyle{fancy}
\pagestyle{fancy}


\begin{abstract}
The progressive integration of electronic converter based Renewable Energy Sources will lead to a reduction of the system synchronous inertia that can endanger the system frequency stability behavior, especially in the case of isolated or poorly interconnected systems, such as the case of Iberian Peninsula system.

This work focuses on the study of the impact that the synchronous inertia reduction  can have on the Iberian system frequency stability for different critical operation scenarios and disturbances.  For this purpose, an equivalent dynamic model of the Iberian system interconnected with the central European network was first developed using the Matlab Simulink graphical programming environment. The parameters of this model are identified using a parameter estimation approach that explores the recorded behavior of the frequency following an outage of a nuclear unit in Spain. This model was used afterwards to perform simulations for different generation outages and multiple operation scenarios between 2020 and 2040.

\end{abstract}



%
\IEEEpeerreviewmaketitle

\section{Introduction}
To meet the decarbonization goals established within European Union many member states have designed CO2 emissions reduction strategies that require large scale integration of renewable power sources (RES) for electricity generation. As a result, significant changes have been promoted in the European Generation Mix over the last ten years, namely a progressive decommissioning of the conventional thermoelectric power stations (Coal, Fuel, Open-cycle gas) and a significant increase  of the RES installed capacity. This transformation of the generation portfolio has occurred at different paces throughout the European Continent, being particularly accelerated in the Portuguese and Spanish case, where an even faster growth of RES is expected for the next 20 years. 

The replacement of coal-fired power plants by  renewable generation, particularly by those connected to the grid through power electronic converters (wind and solar PV), leads to a reduction in the power system synchronous inertia \cite{menos_inercia}. As a consequence of this system inertia reduction, problematic phenomena related with frequency stability, such as large frequency excursions and large $\frac{\partial f}{\partial t}$ variations will become progressively more frequent \cite{aumentodeprobs}. 
The Iberian Peninsula System is a particularly interesting case study, in what regards frequency stability, not only due to the abundance of renewable resources that are used for electricity generation, but also due to its low synchronous interconnection capacity to the rest of the Continental Europe Synchronous Area. Therefore, this work main focus will be the study of the effects of the system synchronous inertia reduction in the frequency stability of Iberian Peninsula System interconnected to the Central Europe System.

The assessment of the system frequency stability is generally performed by carrying out power system dynamic simulations, where one seeks to understand how the system frequency  behaves when facing severe system disturbances, such as generation outages. In general, power system dynamics studies require detailed modelling of all the existing levels of control. However, building a detailed model of the entire Iberian Peninsula and Central European system is quite difficult, not only because to the difficulties in accessing the data regarding grid and generators dynamic models, but also due of the harshness of the process concerned with gathering and assembling all the system parts. 

The two-fold contribution of this paper is to describe a new electromechanical two area equivalent model, including its parameter identification, and to evaluate future stability weaknesses in the Iberian System. With this in mind, an electromechanical two area equivalent model  (Iberian Peninsula and Central Europe) is derived and the parametrization of the model is obtained by exploiting the recorded behaviour of the frequency following the outage of a nuclear power unit of 1 GW. The development of such model allowed performing of a systematic study of the impact of the inertia reduction in the frequency stability due to the increasing penetration of RES. This study used TSO and national projections for the generation mix projections of the generation mix evolution, to evaluate future stability weaknesses of the Iberian System .

In order to tackle with the negative effects for the system frequency stability, that result from the inertia reduction, different TSOs \cite{ffr_eirgrid} \cite{ffr_fin} \cite{9208672} have started to include new ancillary services, such as the provision of Synchronous Inertia, Synthetic Inertia and Fast Frequency Response.An assessment on how synchronous condensers can improve the frequency stability in the Iberian system for some critical operating conditions will also be provided in this paper.

The work presented in this paper is structured in five different chapters. The present chapter  corresponds to the discussion of the theme contextualization and relevancy, as well as of a small literature review. The second chapter,  aims at describing the  developed model structure and its respective parametrization methodology.  Next, in chapter III, the results of the different sensitivity  studies regarding the impact of the system synchronous inertia reduction and the composition of the generation portfolio variation have on the System frequency stability are described.
At last, in chapter IV and V, a brief discussion of possible mitigation measures and a description of the main conclusions are presented.

\vspace{1 cm}

%


\section{Simplified Modelling of Iberian Peninsula system interconnected with
the Central Europe system.}

In order to represent the frequency behavior of the Iberian Peninsula System interconnected with the Central European system, an electromechanical model that includes all the primary control loops regarding frequency response was developed and was later parametrized, by exploring the recorded system frequency response folowing a 1 GW nuclear power plant unit generation outage that took place in Spain in 2011 \cite{evento_ref}.

 \subsection{Model Structure}

\begin{figure}[H]
\centering
\includegraphics[scale=0.55]{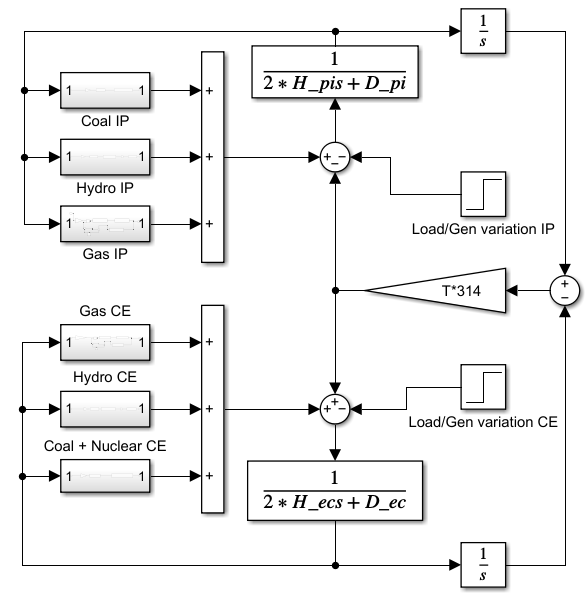}
\caption{Structure of the developed generic model.}
\label{fig:model}
\end{figure}

In the developed electromechanical model, the Iberian Peninsula (IP) and the Central Europea (CE) were considered to be  two distinct but yet interconnected control areas.  Their dynamic behavior was represented using  a  two area control system model, similar to the one presented in \cite{kundur}.  
Since this works focus was on the study of the system frequency stability trough its Nadir and RoCoF behaviors, then, when analyzing large disturbances associated with generation outages it is admissible to represent the dynamic behavior of each generation unit by their primary control loops, without the inclusion of the voltage control loops and the transmission grid effects. Therefore it was considered that the composite response of all the generation units for a certain technology, that takes part in the frequency containment process, can be represented by an equivalent primary control loop. Each control area was then modeled by a swing equation, where the different technologies and  frequency primary  control loops are connected, leading to the block diagram described in  figure \ref{fig:model}. 

As depicted in figure \ref{fig:model}, three equivalent  primary frequency control loops  were considered for each area, with the aim of representing the primary control action of the 4 main synchronous generation technologies in the European generation mix \cite{inert_tip}: Coal fired power plants, Combined Cycle Power plants, Nuclear power plants and Hydro plants.  Since these models should mimic the composite response of multiple machines, then,  simple and generic turbine-governor models were used, namely:

\begin{itemize}
    \item The TGOV1 model, for the case of the nuclear and Coal power plants \cite{pes}.
    \item The GAST model, for the case of the Combined Cycle Power plants and Open Cycle Gas Turbines.
    \item The classic hydroelectric power plant model introduced in \cite{kundur}, for the hydroelectric hydro plants.
\end{itemize} 

It is important to stress that the nuclear power plant primary control was disregarded in the Iberian area, since it was assumed that this baseload units would typically work close to its nominal value, leaving only a very small regulation band available for Frequency Containment Reserve (FCR) provision and that its governor droop setting are typically larger than the ones used in other conventional generation technologies.

The swing equation in each control area represents  the area's loads inertial and damping frequency response and all the area's generators inertial and damping frequency response .  

The existing interconnection between the two control areas, which for the purpose of this study corresponds to the active power flow  between both areas, was modeled in a very simple matter according to the Laplace transform form of equation \ref{eq:tran} .

\begin{equation}
     \Delta P_{1-2}=\Delta \delta_{1-2}\cdot T    
     \label{eq:tran}
\end{equation}

Where:

\noindent T is the well known synchronization coefficient (p.u. GW/ rad).

\noindent $\Delta P_{1-2}$ is the tie-line active power flow variation as a consequence of a frequency variation (p.u. GW).

\noindent $\delta_1$ is the ideal voltage source angular displacement, of the electric equivalent of area 1 (radians).

\noindent $\delta_2$ is the ideal voltage source angular displacement, of the electric equivalent of area 2 (radians).

\noindent $\Delta \delta_{1-2}$=$\Delta \delta_{1}-\Delta \delta_{2}$

\vspace{0.5 cm}

 \subsection{Parameter estimation}
 
If the generic model structure described above is completely parametrized with typical values, then it would not be able to mimic the frequency response behavior of the system. Indeed, this can only be achieved if some of its parameters are carefully calibrated in a way that approximates the model response to the real system behaviour. Since  typical values were used for the time constants of the turbine of the different models \cite{kundur} \cite{pes}, then, the following parameters values were left to be estimated:

\begin{itemize}
    \item The inertia constants of both areas: $H_{CE}$ and $H_{IP}$;
    \item The damping constants of both areas: $D_{CE}$ and $D_{IP}$;
    \item The speed governor droops of all units: $R_{ccgt - CE}$, $R_{hydro - CE}$  $R_{coal \text{  \&}  \text{ } nuclear-CE}$, $R_{ccgt-IP}$, $R_{hydro-IP}$ and $R_{coal-IP}$;
    \item The transient droops of the hydraulic units:  $Rt_{hydro - CE}$  and $Rt_{hydro-IP}$;
    \item The synchronizing torque coefficient $T$;
    \item The speed governor time constants for all technologies $Tg$;
\end{itemize}

A 10 GVA base was used for all the parameters.

\begin{figure}[H]
\centering
\includegraphics[scale=0.27]{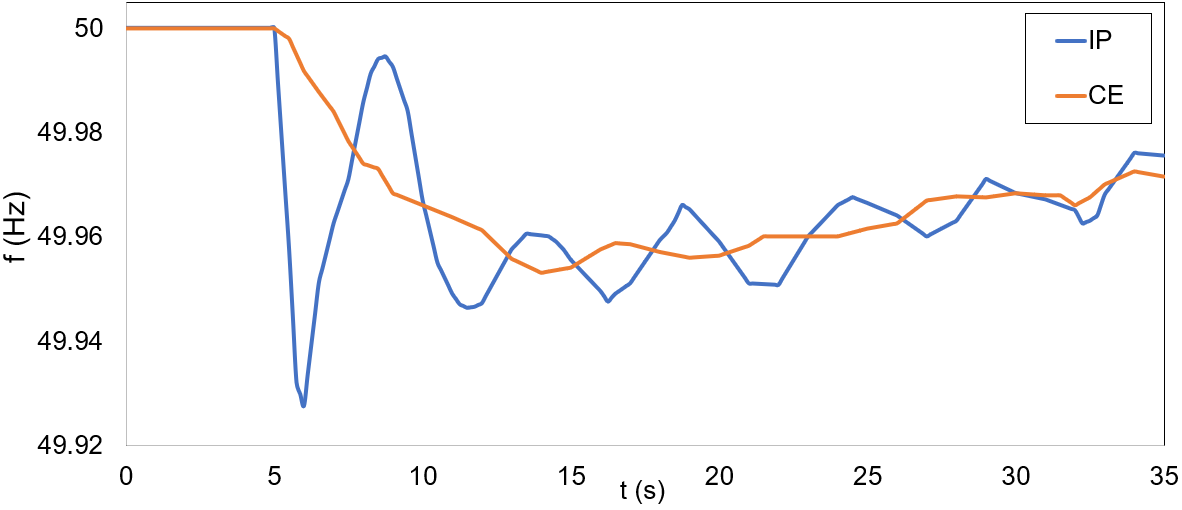}
\caption{Frequency excursion during 2011 plant outage, in Spain \cite{evento_ref}}
\label{fig_model}
\end{figure}

Accordingly, one developed and applied a parameter estimation methodology that tunes the model frequency response to the real system frequency response which was recorded during  a reference disturbance that took place in Spain.\cite{evento_ref}. The mentioned disturbance was the loss of 1 GW of Active Power, due to the Almaraz II Nuclear plant unit disconnection in October 2011. The recorded frequency evolution for the IP and CE area is presented next in figure \ref{fig_model}.

In order to solve the parameter estimation optimization problem, which in this case corresponds to the minimization of  the quadratic error summation between the model frequency response and the real system frequency response (equation \ref{eq:obj}),  a variation of the classical particle swarm algorithm (CDIW-PSO) \cite{pso} was used.

\begin{equation}
\begin{split}
   min:f(y_{IP}(t),\widehat{y}_{IP}(t),y_{CE}(t),\widehat{y}_{CE}(t)) \\ =\sum_{t=1}^{NT}(y_{IP}(t)-\widehat{y}_{IP}(t))^2+\sum_{t=1}^{NT}(y_{CE}(t)-\widehat{y}_{CE}(t))^2
   \end{split}
   \label{eq:obj}
\end{equation}

Where:

\noindent $\widehat{y}$ is the model frequency response.

\noindent $y$ is the real system frequency response.

As it can be seen in equation \ref{eq:obj}, the objective function depends on the model frequency response, which is obtained via simulation on the Matlab Simulink platform. This means that in the evaluation phase of the algorithm, which is performed in every iteration of the PSO, a numeric integration simulation is carried out.  Naturally, this means that the parameters under analysis cannot take any value, since the numeric integration algorithm would not  converge, what would lead to a premature stoppage of the PSO. Multiple constraints were included in the CDIW-PSO algorithm to obtain the desired output. These were modelled as parameter values restrictions, when its  purpose is to ensure convergence in the Simulink and as penalty functions, when its purpose it to grant that the final parameters values are  plausible.

\begin{figure}[H]
\centering
\includegraphics[scale=0.3]{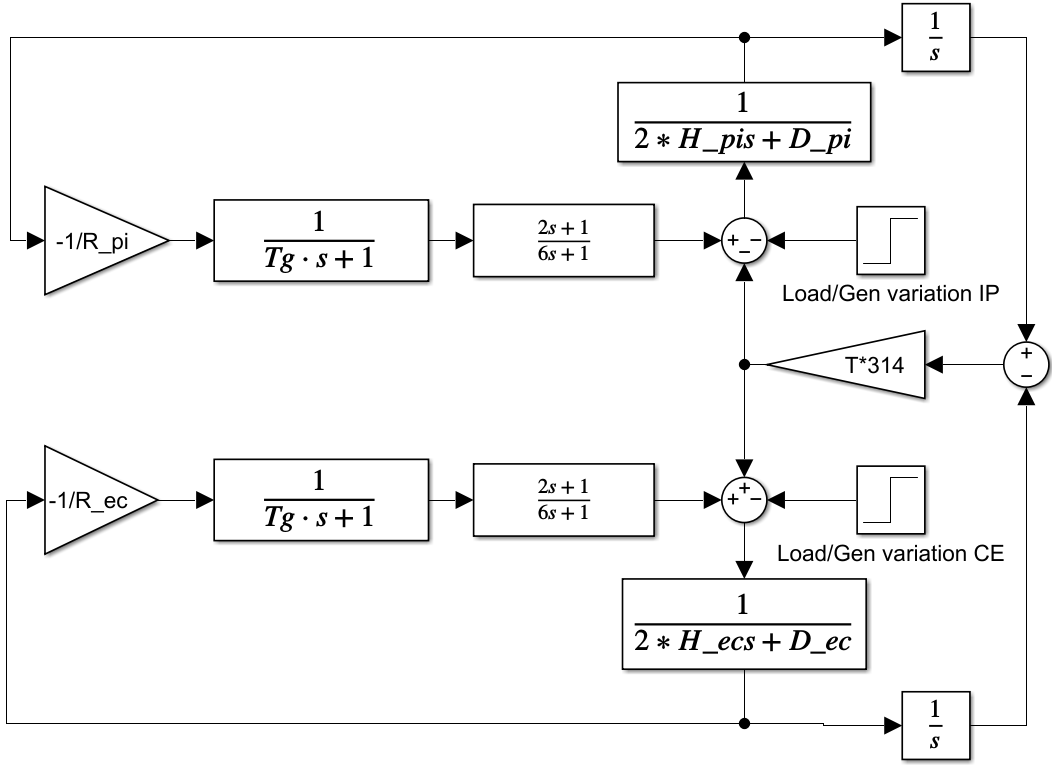}
\caption{Structure of the intermediate generic model.}
\label{fig:model_min}
\end{figure}

The CDIW-PSO was then implemented in the Matlab programming platform with the aim of calibrating the parameters of the model presented in figure \ref{fig:model}. However, given the large number of parameters to estimate (12), an intermediate process that corresponds to the estimation of the parameters of the model depicted in figure \ref{fig:model_min} was introduced. In this smaller model, each area equivalent primary control loop is modelled using a simple TGOV1, allowing for a reduction of the number of parameters to estimate from 14 to 8, which translates into a much faster computation time.

After the above model parameters are estimated their values can be used as a starting point for the calibration of the complete model parameters. When doing so one must allocate each area equivalent droop value through the different equivalent technologies, something that was performed using historical data of the generation mix in homologous periods to the one of the incident \cite{ree_data} \cite{ren_data}.

\begin{figure}[H]
    \centering
    \includegraphics[scale=0.285]{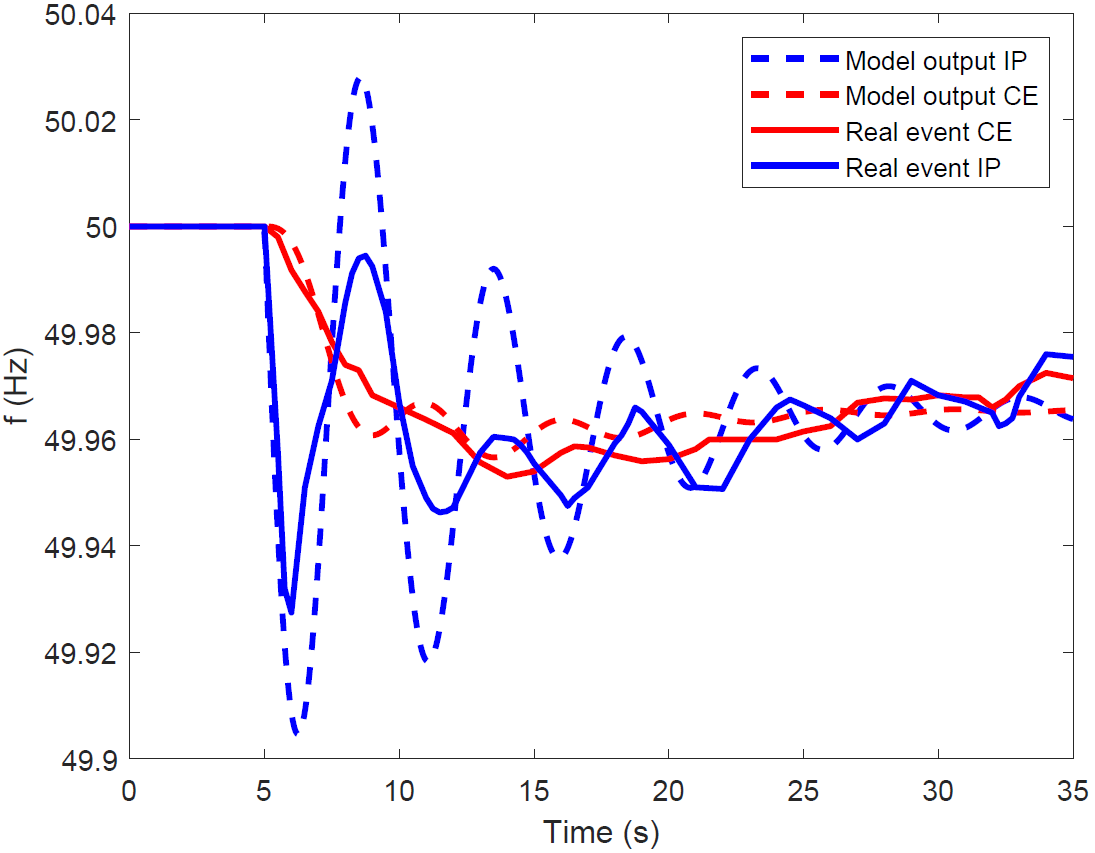}
\caption{Real system and Calibrated Model frequency response to 1 GW generation outage. }
    \label{fig:comp}
\end{figure}

\begin{figure}[H]
    \centering
    \includegraphics[scale=0.37]{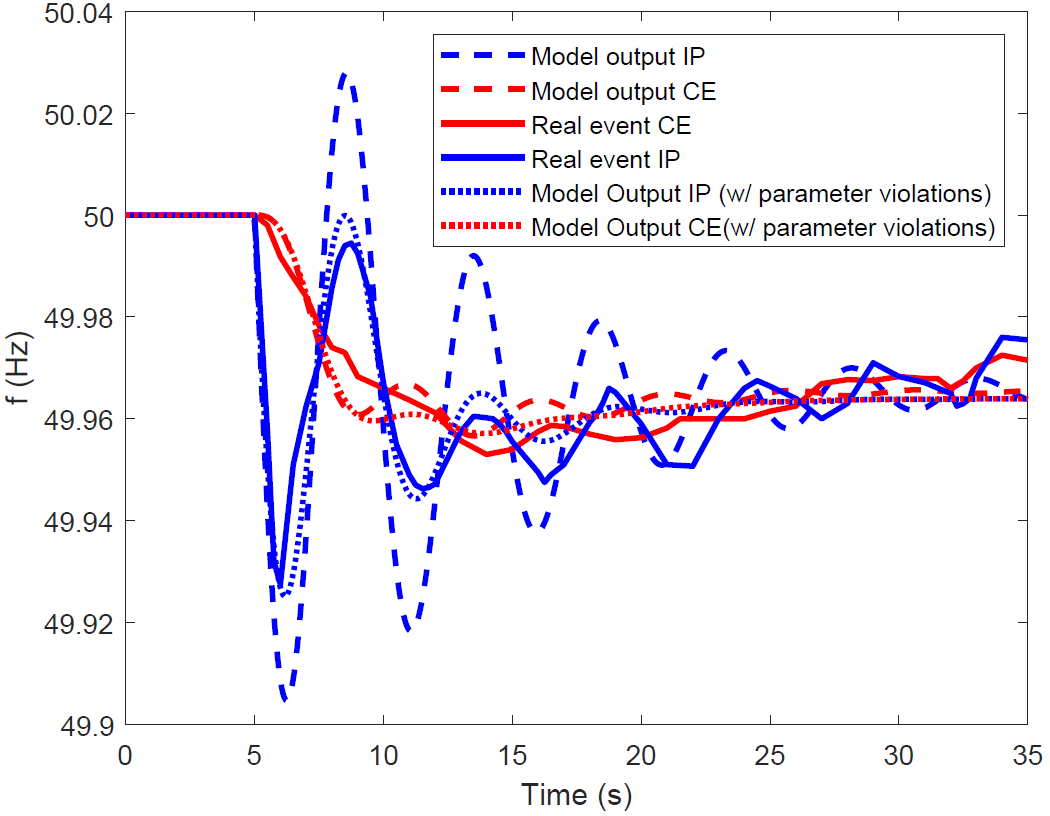}
   \caption{Example of the model frequency response for other parameter value combinations that violated defined constraints. }
    \label{fig:comp_2}
\end{figure}

 At the end of the parameter estimation process successful results were achieved, since  the estimated values are not only plausible, in the sense that no negative values  or inexplicable relations between variables resulted from the estimation process, but they also seem to be in conformity with the real operation conditions of the Iberian Peninsula system interconnected with the central European system.  In addition, the parametrized model frequency response also showed adequate adherence to the real system frequency response following the Almaraz II generation outage, as depicted in figure \ref{fig:comp}. It is worth mentioning that a better adherence between the two curves was achieved with different sets of parameters, as shown in figure \ref{fig:comp_2}, however these were discarded since at least one of the parameters values violated the reasonable limits.

When comparing the different curves of  figure \ref{fig:comp_2} no relevant differences could be seen in the Rocof value, for measuring windows smaller or equal to 500 ms, however, the Nadir value varied in a range of 25 mHz. As mentioned before, the set of parameters that leads to the large dashed line were chosen over the ones that lead to the small dashed curve, first because of its compliance with the parameter constraints and secondly  because it would lead to a worst case hypothesis for the Nadir value without introducing any significant error.

 \subsection{Scenario representation in the model}
 It is evident that when using an incident that took place in a specific moment of time, namely
in October 2011, one can not expect that the estimation resulting parameters, like the inertia and the equivalent governor droop setting, stay the same throughout the years or even throughout that specific year. Nevertheless, these parameters characterize the system in a moment of time $o$, therefore if enough information about the generation mix is available, then, they can be used as a reference or starting point for the deduction of the parameter values that may represent the system behavior in other moment $t$. With this in mind, different model parameters  can be adjusted to represent the system state in an instant $t$, as follow:

    \begin{itemize}
        \item Equivalent speed droop setting of technology $i$ . 
    \end{itemize}
      \begin{equation}
R_{ti}=R_{oi} \cdot \frac{Pg_{oi}}{Pg_{ti}}
\end{equation}
 Where:
 \noindent $Pg_{oi}$ is the power being generated by technology  $i$ in the moment of the Almaraz II outage.
 \noindent $Pg_{ti}$ is the power being generated by technology  $i$ in instant $t$.
 
    \begin{itemize}
        \item Damping constant.
    \end{itemize}
    
      \begin{equation}
D_{t}=D_{o} \cdot \frac{Pd_{t}}{Pd_{o}}
\end{equation}

  Where:
 \noindent $Pd_{0}$ is the system  load in the moment of the Almaraz II outage.
 \noindent $Pg_{t}$  is the system  load  in instant $t$.
    
    \begin{itemize}
        \item Inertia constant. 
    \end{itemize}
    
       Even though the Inertia Constant (H) was modelled and estimated with the assumption that it represents the whole area inertial response, when adjusting it for different scenarios this variable must be first split into parts.
    \begin{equation}
    H_{o}=\sum_{i=1}^5 H_{oi} + Hload_{o}
    \label{eq:inercia}
    \end{equation}
    
    In equation \ref{eq:inercia} $H_{oi}$ represents the equivalent inertia constant of all the machines of technology $i$ in the reference instance and it was calculated using $Pg_{to}$ and typical inertia constant values for each technology \cite{inert_tip}. The other part $Hload_{o}$ represents the load inertial response and it can be calculated as the difference between $H_{o}$ and the generation inertia. The generation parcel of the inertia can then be adjusted as follows:
    
    \begin{equation}
        H_{ti}=H_{oi} \cdot \frac{Pg_{ti}}{Pg_{oi}}
    \end{equation}
    

Although the load inertia parcel can  be adjusted to the load size, in this work, one made the simplification of considering it constant.

 \section{Sensitivity Studies regarding inertia variation and composition of the generation portfolio}
 
  Having a simulation platform that is conveniently calibrated with data from a real system disturbance and whose parameters can be adjusted to represent different generation scenarios, then, one could start to address the focal point of this work - the assessment of the frequency stability behavior for the future evolution of the Iberian System. With this purpose in mind, multiple simulations of the reference disturbance (outage of a 1 GW generation in the IP)  were performed in the developed model, considering different load and generation scenarios. These scenarios, that  are detailed in the case study section, are an attempt to represent the changes that are expected to take place in the Iberian system, in the following 20 years from now.
  
The performance of these simulations will allow the identification of possible high-risk scenarios to the system frequency stability, before they actually happen. Preventive measures can then be antecipated and later taken to avoid the materialization of those scenarios in the future.

 \subsection{Case study scenarios}
 
The simulation of the reference disturbance was performed for three different scenarios: the Monthly Average scenario, the Winter Valley scenario and the Peak PV hour scenario.  Each of these scenarios is characterized by different load and generation mix conditions, that change throughout the analysis time horizon (20 years).

\subsubsection{ Monthly Average scenario}

\begin{figure}[H]
     \centering
     \includegraphics[scale=0.42]{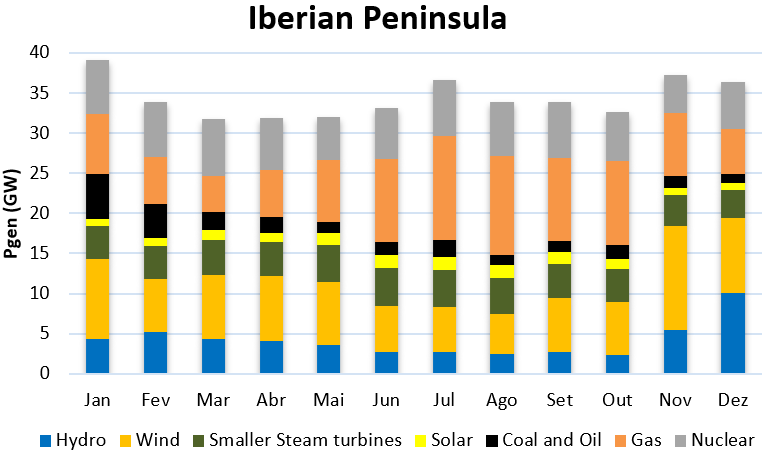}
     \caption{Starting point for the Monthly Average scenario (Iberian Peninsula). }
     \label{fig:incial_average_1}
 \end{figure}

\begin{figure}[H]
     \centering
     \includegraphics[scale=0.506]{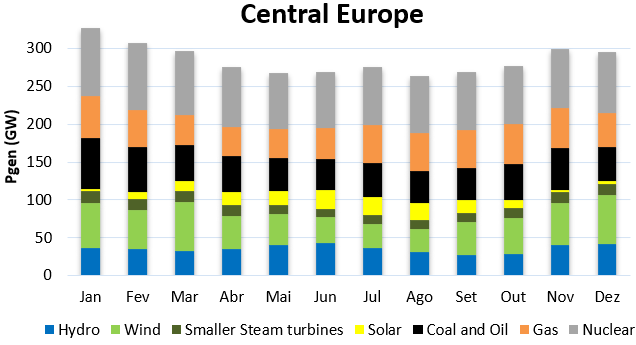}
     \caption{Starting point for the Monthly Average scenario (Central Europe). }
     \label{fig:incial_average_2}
 \end{figure}

In this annual scenario, it was considered that each month could be characterized by a single dispatch scenario, that reflects the average hourly production of each technology type in that month. As starting point for the Iberian area, data from 2019 ,that is available in the website of both the Portuguese and Spanish TSOs, was used and for the Central Europe case, one resorted to Eurostats \cite{ren_data} \cite{ree_data} . The achieved dispatch scenarios for the different months, in the two control areas, are depicted next in figures \ref{fig:incial_average_1} and \ref{fig:incial_average_2}.
 
 Notice here that the generation category identified in the above figures as Smaller Steam Turbines includes Biomass, Cogeneration and Solar Thermal units.
 
 In what regards the evolution of the Iberian Peninsula generation mix until 2040, the following assumptions were made:

 \begin{itemize}
   \item	All coal fired power plants are to be decommissioned until the end of 2030 and all nuclear plants are to be decommissioned until the end of 2035.
\item	The share of solar PV generation in the annual generation mix will grow from 4 to 24 $\%$.
\item	The share of wind generation in  the annual generation mix will grow from 22 to 29 $\%$.
\item	The share of Small steam turbines is considered constant throughout the analysis horizon.  
\item	The share of hydro plants is considered constant throughout the analysis horizon. 
\item	Natural gas power plants (CCGT) are used mainly to keep security of supply, therefore its generation presence will depend on the amount of load that is let unfed by the other technologies.

 \end{itemize}
 
 On the other hand for the Central European area it was considered that the generation mix would evolve into the 2040 Distributed Energy scenario as described in the TYNDP 2020 \cite{tyndp2020}.
 
 It's important to underline that this first scenario does not allow for a true identification of the most critical hours of operation, since it reproduces the average conditions of the system during a specific month of each year. Nevertheless, its study allows for the recognition of the path that the system its taking, in what regards frequency stability , and can also give good indications of the periods of the year where the system synchronous inertia volume will be shorter.

 \subsubsection{Winter Valley scenario}
 
Contrarily to the Monthly average scenario, the Winter Valley scenario does not have monthly discrimination, therefore only a single dispatch scenario,  that reflects the expected system conditions  during the winter nights,  is considered for every year.

As a starting point for the Iberian Peninsula area,  2019 data from \cite{ren_data} \cite{ree_data} was once again used. It was also assumed that   each technology type dispatch could be represented by its average value during the periods between 22 p.m and 5 a.m, of the  winter nights (characterized by large wind power generation). Additionally,  the  evolution of the generation portfolio throughout the years, in these intervals, was considered to unfold as follows:

\begin{itemize}
    \item	All coal plants are to be decommissioned until the end of 2030 and all nuclear plants are to be decommissioned until the end of 2035.
\item	The share of wind generation will grow from 29 to 83 $\%$.
\item	The share of Small steam turbines  is considered constant throughout the analysis horizon.
\item	In what regards the share of hydro and  combined cycle gas turbine power plants, it was considered that both would experience a reduction due to the growth of wind generation. Nevertheless, an equivalent hydro power plant of about 1 GVA was still considered to be online in 2040, to fulfill reserve needs.

\end{itemize}

\begin{figure}[H]
    \centering
    \includegraphics[scale=0.47]{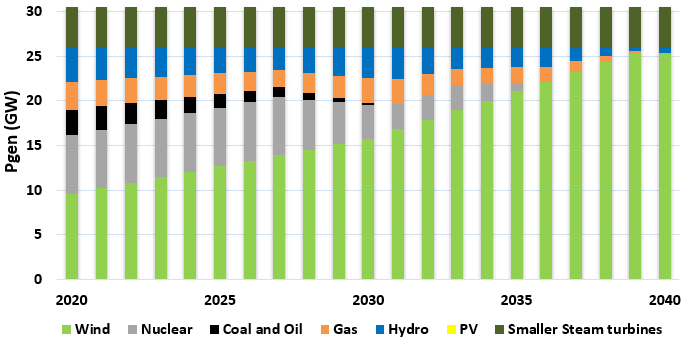}
    \caption{Evolution of the generation mix in the IP area, for the Winter Valley scenario.}
    \label{fig:winter_evoll}
\end{figure}

All things considered,  a generation mix  evolution that  is displayed next in figure \ref{fig:winter_evoll} was then reached.

\begin{figure}[H]
    \centering
    \includegraphics[scale=0.52]{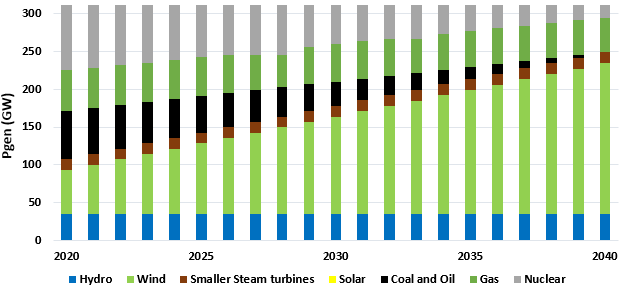}
    \caption{Generation mix evolution in the CE area, for the Winter Valley scenario.}
    \label{fig:winter_evoll_ec}
\end{figure}

In what regards the Central Europe a more simplistic scenarization was considered, namely an adjustment of the monthly average scenario that corresponds to the night load value for January. The adopted generation mix evolution can be seen in figure \ref{fig:winter_evoll_ec}.

  \subsubsection{ Peak PV hour scenario}
  
Given the foreseen growth  of the PV generation installed capacity in the Iberian Peninsula, we already witness a reduction of the synchronous generation presence in the hours of large PV production (11 am to 3 pm). Since this growth tendency is expected to continue,  these will soon become one of the periods where the system inertia is smaller, and therefore its study is inevitable.  The peak PV hour scenario, as the winter night scenario, is composed out of  21  dispatch scenarios (one for each year under analysis ranging from 2020 to 2040),  that reflect the expected system operating conditions  during the hours of higher penetration of PV generation.

 Similarly to the winter night scenario,  the starting point for the Iberian Peninsula  generation mix was considered to be the distribution that is averagely found during the peak PV hours of the summer . In other words,  the assumption that  each technology type dispatch could be represented by its average value during the periods between 11 a.m and 3 p.m in the  summer days was made. Additionally, it was considered that the generation portfolio evolution throughout the years, in these specific periods, unfolds as follows:

\begin{itemize}
\item	All coal plants are to be decommissioned until the end of 2030 and all nuclear plants are to be decommissioned until the end of 2035.
\item	The share of PV generation will grow from 9 to 66 $\%$.
\item	The share of Small steam turbines and wind generation  is considered constant throughout the analysis horizon.
\item	In what regards the share of hydro and  combined cycle gas turbine power plants, it was considered that both would experience a dispatch reduction due to the growth of wind and solar generation. Nevertheless, an equivalent CCGT power plant of about 1 GVA was still considered to be online in 2040, to fulfill reserve needs.

\end{itemize}

\begin{figure}[H]
    \centering
    \includegraphics[scale=0.37]{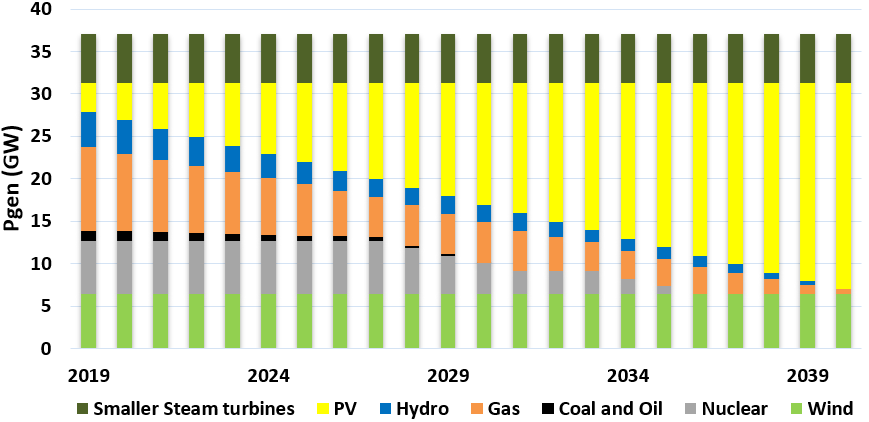}
    \caption{Generation mix evolution in the IP area, for the Peak PV hour scenario.}
    \label{fig:pv_evolui_x}
\end{figure}

  All things considered,  the generation mix  evolution that  is displayed in figure \ref{fig:pv_evolui_x} was reached.
  
\begin{figure}[H]
    \centering
    \includegraphics[scale=0.52]{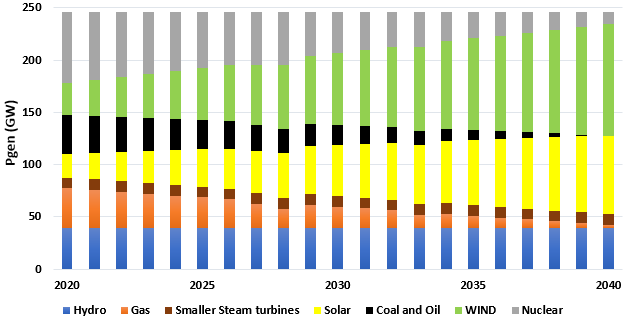}
\caption{Generation mix evolution in the CE area, for the Peak PV hour scenario.}
    \label{fig:peak_pv_ec}
\end{figure}

Again, for the Central Europe area  small adjustments were made to the monthly average scenario  that corresponds to June, in order to represent the CE system conditions in the hours between 11 p.m and 3 a.m. The achieved generation mix evolution is depicted next in figure \ref{fig:peak_pv_ec}.
 
 \subsection{Simulation results}

\begin{figure}[H]
    \centering
    \includegraphics[scale=0.29]{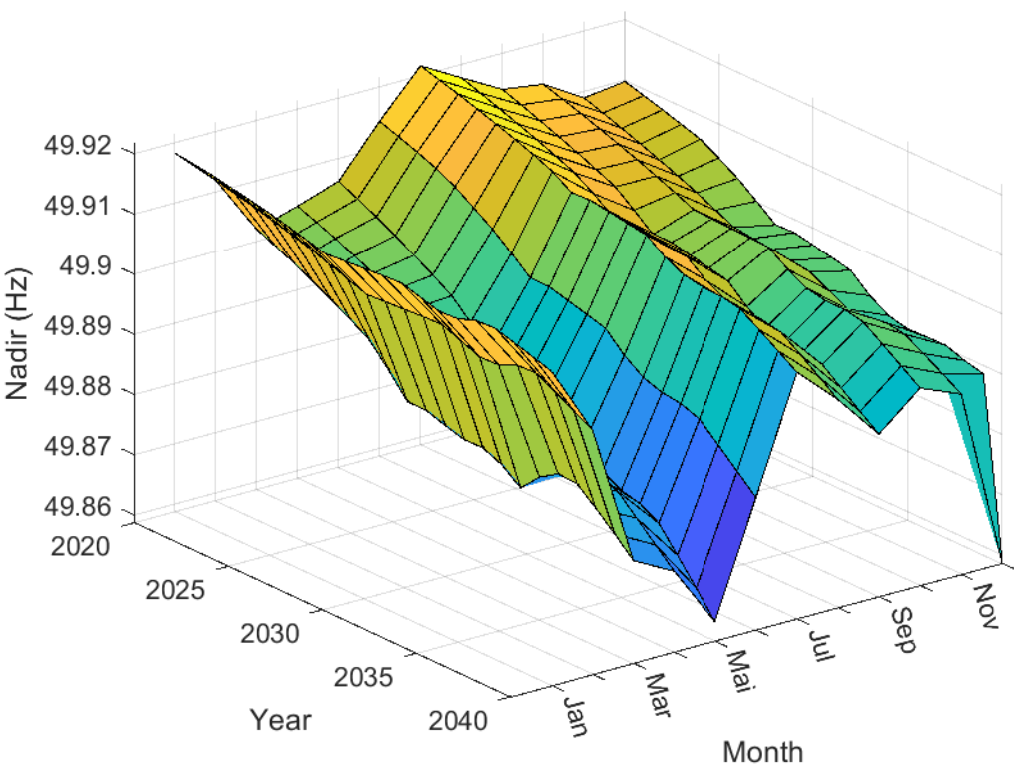} 
  \includegraphics[scale=0.29]{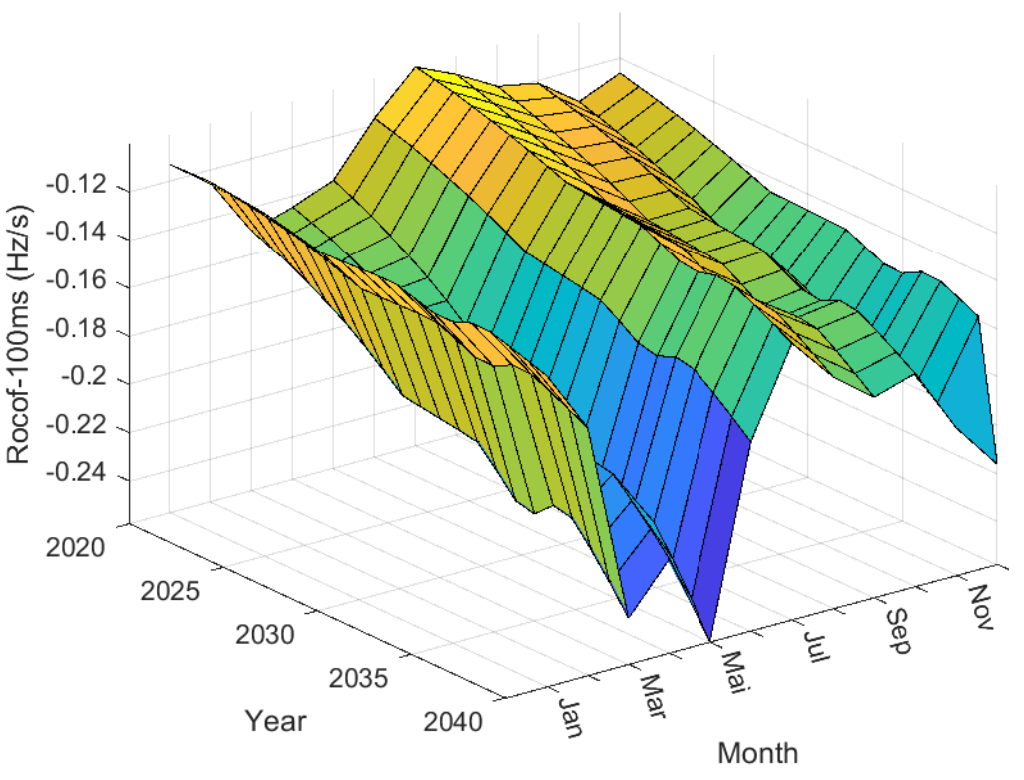} 
   \caption{IP's Nadir and RoCoF evolution, in the Monthly Average scenario.}
         \label{fig:3d_average}
 \end{figure}

The approach described before allows then to adjust the dynamic model parameters to the different dispatch scenarios.
Next,  multiple simulation of the reference  disturbance, one for each
year/month,   are performed and the respective IP area frequency Nadir and Rocof (for a 100 ms sliding window) are recorded. The results achieved for the  Monthly Average scenario are presented next in figure \ref{fig:3d_average}.

 Before commenting on the Rocof evolution itself, it is important to underline that the Rocof values that are here plotted corresponds to a calculation for a 100-ms sliding window. Even though the operational limit and withstand capability Rocof values are usually determined in a 500-ms time window,  a 100 ms time window was here chosen since its closer to provide a better indication of the scale and range of the challenges \cite{last_entsoe} the system will have to face. 
 
The Nadir and Rocof evolution presented in the above figures shows that even in average scenarios both the Rocof and frequency deviation tend to double in a time-span of 20 years. Something  that suggests that in a majority of the dispatch scenarios the system will face increasingly complex operating conditions in what regards frequency stability.  

\begin{figure}[H]
    \centering
    {{\includegraphics[scale=0.28]{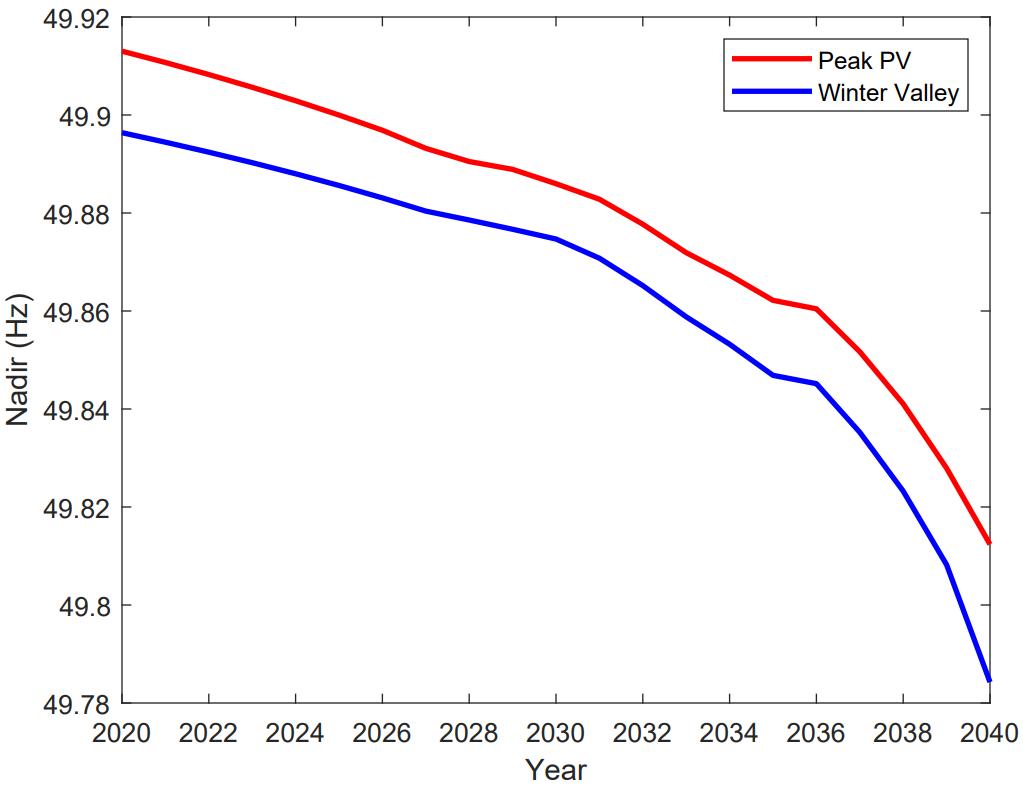} }}
    \qquad
  {{\includegraphics[scale=0.28]{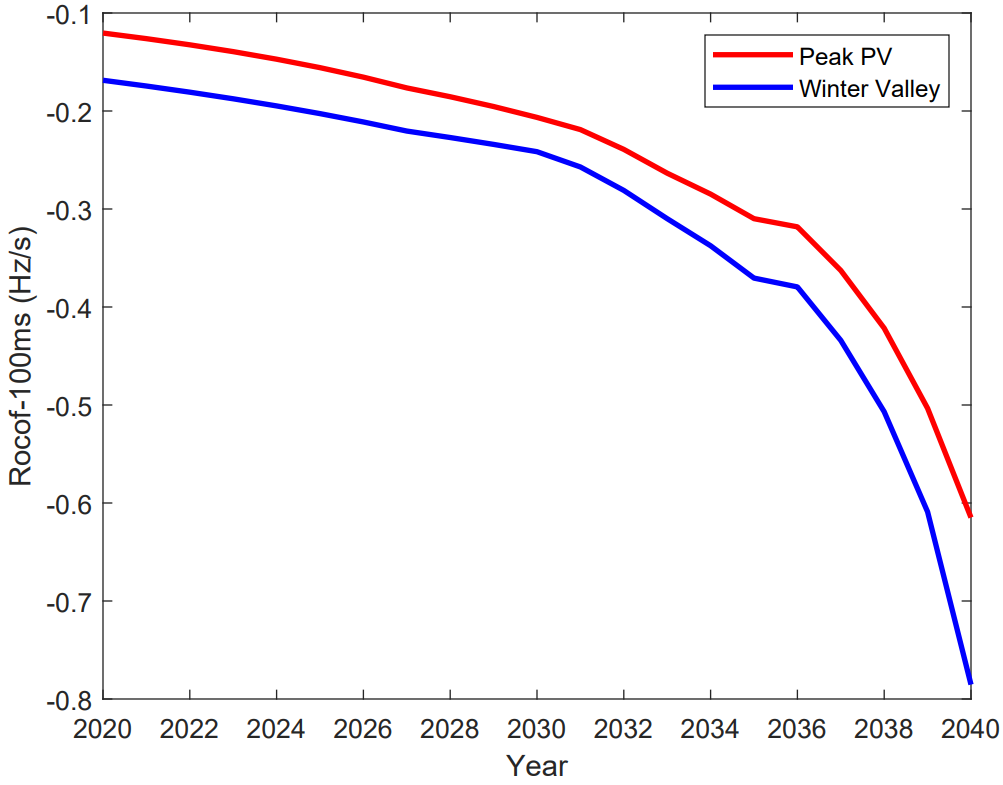} }}
   \caption{IP's Nadir and RoCoF evolution, for the Peak PV hour scenario and winter nights scenario, with different cogeneration capacities.}
         \label{fig:spec}
 \end{figure}

With this in mind, one will now step into the analysis of specific periods where the system  has typically lower synchronous generation presence, since this translates into smaller system synchronous inertia.  The scenarios that embody these characteristics were already introduced in the prior section, Winter Valley and Peak PV hours. Again, a 1 GW outage taking place in the PI area was simulated for every dispatch scenario and the corresponding Nadir and Rocof values were recorded. These results are now presented in graphical form in figure \ref{fig:spec}.

 Based on the results of figure \ref{fig:spec}, it seems to be clear that if a generation outage of 1 GW takes place during the hours of higher PV penetration or during the Winter night valley hours, in the years after 2030, the system will experience increasingly larger Rocofs. Despite the frequency Nadir being way above of its boundary condition for load shedding (49.2 Hz), the RocoF is heading towards the 1 Hz/s limit, which leaves the system in a state where the frequency value changes rapidly. 
 
It is also worth mentioning that both the Nadir and RoCoF curve end up reaching lower values  in the Winter Valley hours scenario than in the Peak PV hour scenario . In what concerns the RoCoF, this is massively justified because of the presence of synchronous solar thermal units in the Peak during day light hours scenario and due to the difference between the inertia constant  of hydraulic  and CCGT units. On the other hand, regarding the frequency Nadir, not only the effect of the different inertia is present but also the inherent worst frequency support provided by hydraulic units \cite{kundur}, namely the effect of the water starting time. All these circumstances make  future winter night scenarios vulnerable regarding frequency stability.

Nevertheless, the fact is that even in high renewable penetration scenarios, such as the ones under analysis, the Rocof did not surpass 1 Hz/s. In both scenarios there  is still a  significant synchronous machine’s presence, namely of different types of small scale steam turbines that are spread throughout the grid. These types of generation units that easily amount to 4 or 5 GW in the day to day mix, result from cogeneration and biomass units, showed the capacity to provide  a sufficient inertial response that avoids the surpassing the 1 Hz/s Rocof limit. 

\begin{figure}[H]
    \centering
    {{\includegraphics[scale=0.3]{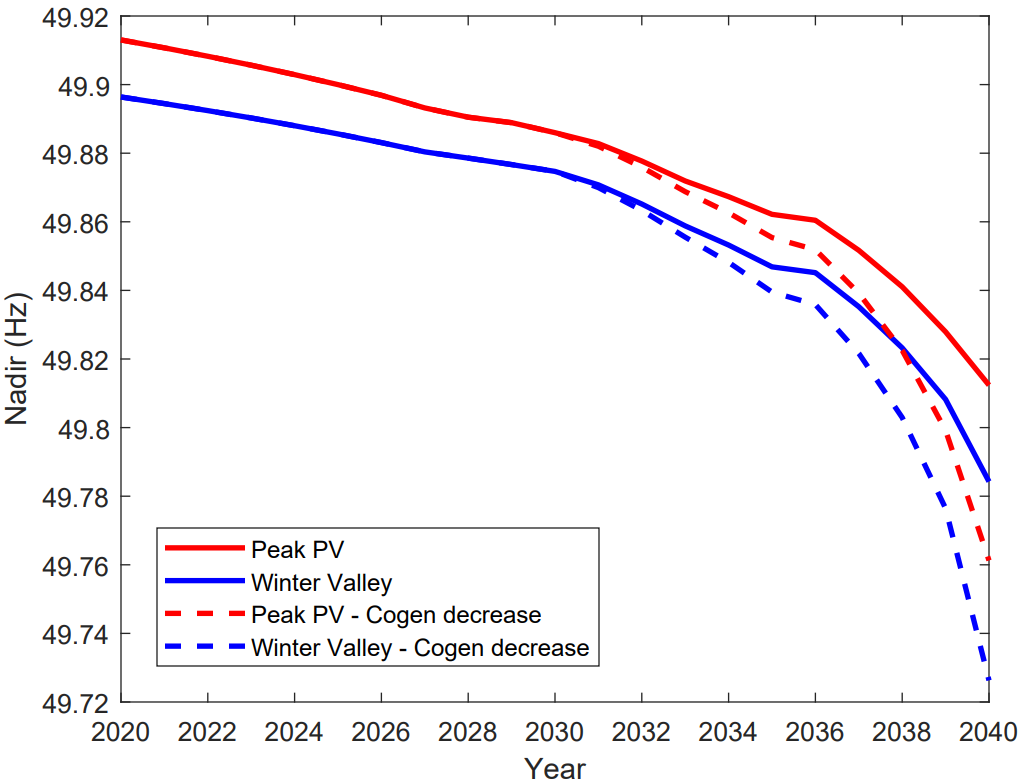} }}
    \qquad
  {{\includegraphics[scale=0.3]{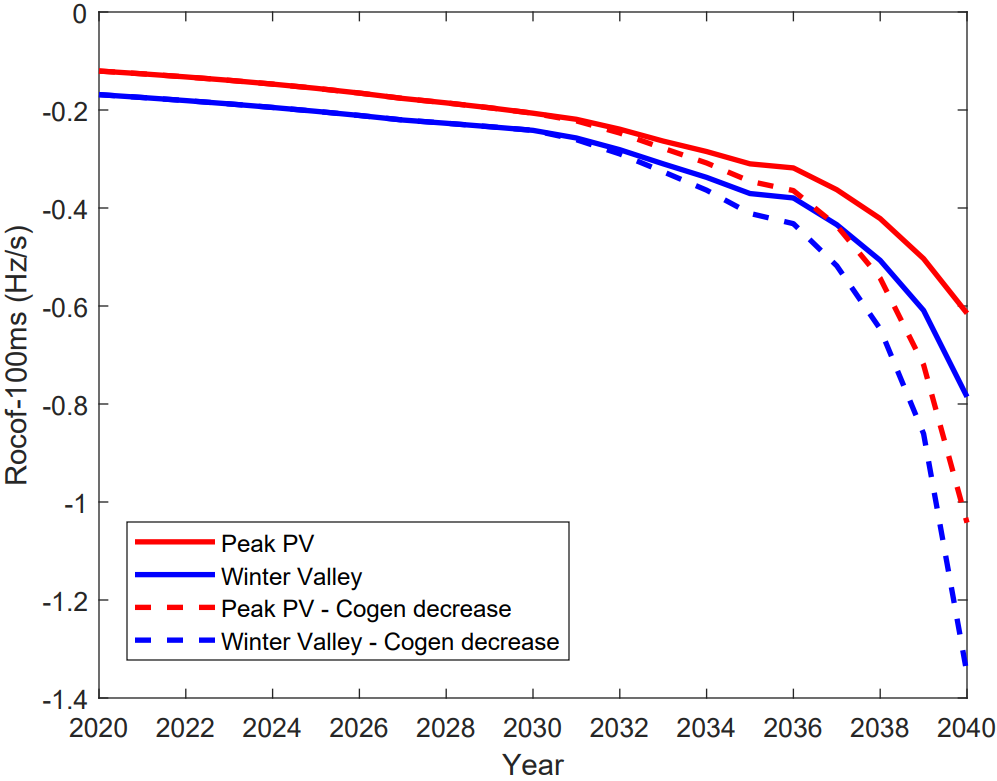} }}
   \caption{IP's Nadir and RoCoF evolution, for the peak PV hour scenario and winter nights scenario.}
         \label{fig:spec_final}
 \end{figure}

Even though the presence of this type of generation was assumed to be constant throughout the years, that may not be the case. As explained before, in the small steam turbine generation category different types of generation plants were included: biomass, solar thermal and cogeneration plants. In the Iberian Peninsula the large majority of this generation category is associated with industry cogeneration plants, that use as a fuel natural gas. Given the plans to reduce and abandon natural gas, it may be possible that different types of heat sources are found and some of these facilities end up getting decommissioned. Therefore, to test the impacts of this hypothesis one last study was performed, where the small steam turbine presence in the mix, for both scenarios, starts reducing in the years after 2030 and tends to half of its initial value in 2040.  The results that concern the simulation of the 1 GW outage, in these modified scenarios, is presented in figure \ref{fig:spec_final}.

A swift analysis of the dashed line curves, presented in  figure \ref{fig:spec_final}, shows that in scenarios with high renewable penetration, the amount of small steam turbines that are connected to the system may be a decisive factor in what regards to the violation of the Rocof limit.

 In fact, independent of the scenario, the 1 Hz/s limit was surpassed in the year 2040, when considering a reduction in half of the steam turbine presence. This is a problem, especially in events like the one being analyzed that due to its size can end up having a cascade of effects and leading to other system outages \cite{r1_zonaecsa} \cite{last_entsoe}  .

This small assessment clearly shows  that the small steam turbines that are spread throughout the Iberian system may be an important asset in what regards the system frequency stability.

\section{Mitigating Measures}

The analysis presented in the prior chapter validates what has already been stated by ENTSO-E and some of the Nordic TSOs, that in low system inertia situations (high converter based RES penetration) the system frequency stability may be compromised, especially in systems with low interconnection capacity. That being the case,  supplementary control systems and other corrective measures need to be envisioned and then implemented to ensure that the future Iberian  System remains secure from the dynamic point of view.
Examples of these measures are:

\begin{itemize}
  \item	The provision of synthetic inertia by converter based generation, which allows for a response proportional to the $df/dt$,  that showed to have impact on the Nadir and Rocof mitigation \cite{f_inertia_2}. Nevertheless, in  scenarios of markedly small inertia its effect on the Rocof seems to be limited by the time delay introduced by the frequency measurement process. Something that clearly distinguishes these type of service from real synchronous inertia, which is an inherent and automatic response of machines.
\item	The supplementation of the system synchronous inertia, using synchronous condensers or power station synchronous generators when they are not being
operated as generator (operation as synchronous motor).  What can still be a viable option, since this solution provides multiple other needed services to the grid, such as  voltage control, reactive power control and short-circuit currents.
\item	The use of converter generation with grid forming capabilities.
\end{itemize}

\subsection{Effect of the synchronous condensers introduction in the frequency stability of the Iberian System}

Independently of the measures that are chosen, TSOs should avoid using solutions like definition of must-run units since this may lead to curtailment of renewable generation which should be avoided as much as possible.

\begin{figure}[H]
    \centering
    {{\includegraphics[scale=0.3]{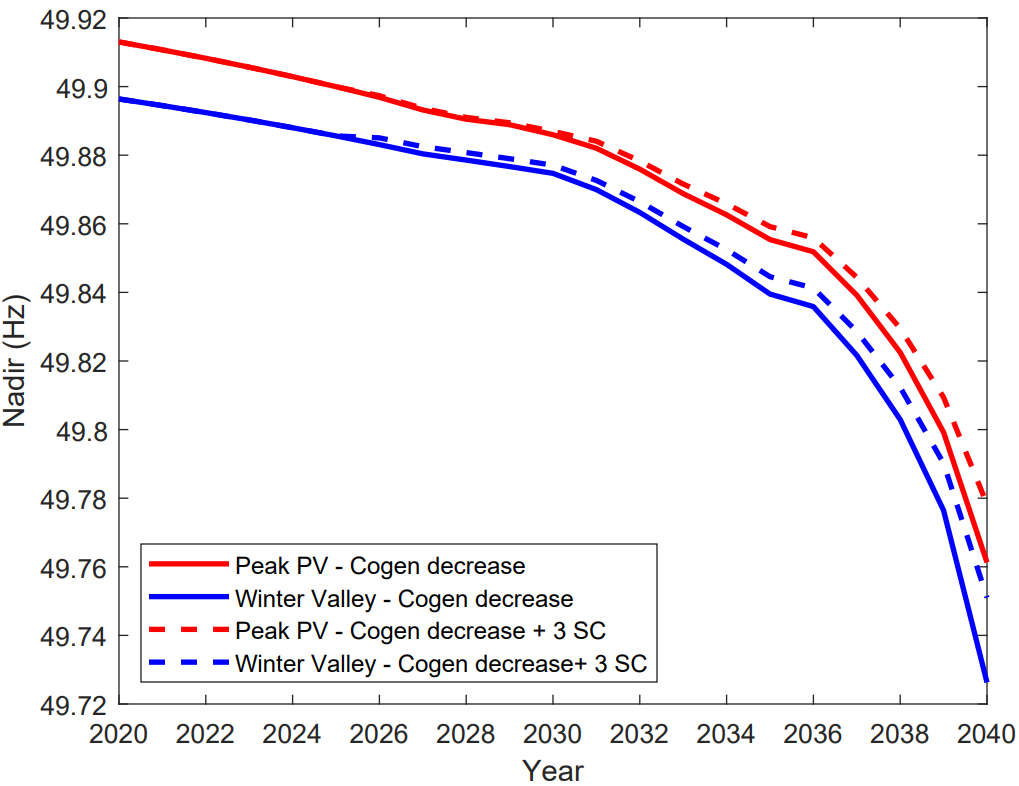} }}
   \caption{IP's Nadir evolution, when including 3 SC from 2025 onward.}
         \label{fig:melhorias_compensador_r}
 \end{figure}

\begin{figure}[H]
    \centering
  {{\includegraphics[scale=0.3]{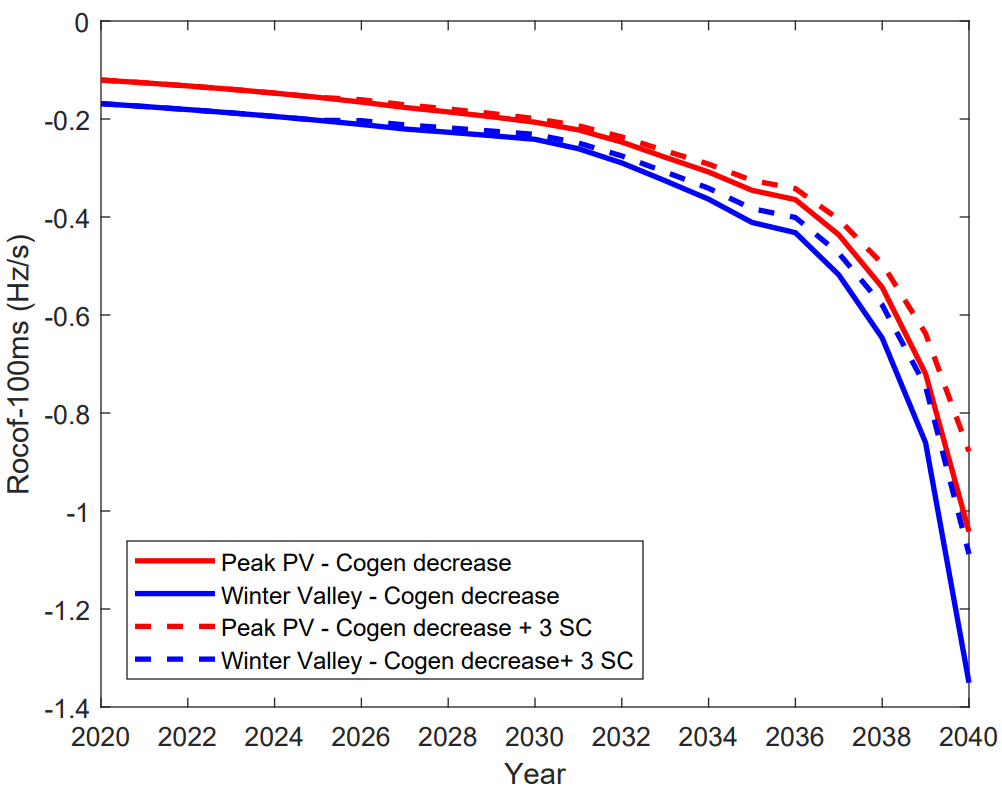} }}
   \caption{IP's RoCoF evolution, when including 3 SC from 2025 onward.}
         \label{fig:melhorias_compensador}
 \end{figure}

For illustrative purposes, a simplified analysis of the impact of the introduction of synchronous condensers in the system stability indexes is presented in figures \ref{fig:melhorias_compensador_r} and \ref{fig:melhorias_compensador}. In this assessment, 3 synchronous condensers were considered to be online from 2025 onward. Each SC was considered to have a inertia constant equal to the average inertia constant of the recently decommissioned groups in  Sines and Pego Coal fired power stations. The scenario where a cogeneration presence  is reduced from the  2030 onwards is used, since it was the one where the largest Rocof values were identified.

As it can be seen in the two prior figures, independently of the scenario, the introduction of synchronous condensers in the system has a clear positive impact on the frequency Nadir and RoCoF 100. It is also worth mentioning that for the specific case of the winter valley, the RoCoF 100 value in 2040  is below the security threshold of 1 Hz/s.

\section{Conclusion}

In this work, multiple studies were carried out, in order to access the ability of the future Iberian System to maintain frequency stability in different load and generation scenarios. For this purpose a simulation platform/model was developed being its parameters conveniently calibrated with data from a real system disturbance. Different relevant simulations were performed for different operating conditions in order to identify the RoCoF and the frequency Nadir, which were later used to compare the system ability to respond to large generation outages.

It is now clear that the transition to a generation portfolio mainly composed out of solar and wind energy is  taking the Iberian system into a state where the maximum frequency deviations and the RoCoF absolute values will be significantly larger than the ones found today. When considering the reference disturbance (1 GW outage), it will not be a surprise if by 2040 a 0.2 Hz decrease happens in the Nadir and that the RoCoF values tend to be 6 times larger or more, depending on the generation scenario. 

If the 1 Hz/s Rocof limit is surpassed, as identified,  the system may enter in a alert zone, where is actually hard to predict if a cascade of events will happen. Indeed, in these circumstances two especially critical events can take place: 1) the triggering of the RoCoF protection of some of the grid synchronous generators; 2) the triggering of the frequency protection of old solar/wind power  plants and of different co generation plants \cite{sep}. Both  events contribute to the enlargement of the disturbance size and result in deeper frequency excursions  that will leave the system in a state where emergency measures, such has load shed, will probably be needed although undesirable.
 
To conclude, even though the frequency Nadir does not seem to be a concern given the fact that its simulation  value was always far from the limit defined by the European regulations, the foreseen $df/dt$ may result in a non secure system operation state. Therefore, it is important to start to prepare the system with supplementary control solutions that provide additional frequency support, such as synchronous condensers and synthetic inertia, preventing in this way frequency instability namely in very high renewable penetration scenarios.






%

\bibliographystyle{IEEEtran}
\bibliography{ref}

\end{document}